\documentclass[preprint, proceedings]{rmaa}

\usepackage{natbib}
\usepackage{paralist}

\newcommand{\tacc}{\tau_{\rm a}}
\newcommand{\tmag}{\tau_{\rm m}}

\newcommand{\mdotacc}{\dot M_{\rm a}}
\newcommand{\gammacrit}{\gamma_{\rm c}}

\newcommand{\rco}{R_{\rm co}}
\newcommand{\rt}{R_{\rm t}}

\newcommand{\rout}{R_{\rm out}}

\newcommand{\spineq}{T_{\rm eq}}
 
\SetYear{2004}
\SetConfTitle{Gravitational Collapse: From Massive Stars to Planets}

\title{The Spin History of Protostars: Disk Locking, Revisited}

\author{Sean Matt\altaffilmark{1,2} 
    and Ralph E. Pudritz\altaffilmark{1}}

\altaffiltext{1}{McMaster University, Canada.}
\altaffiltext{2}{CITA National Fellow.}

\fulladdresses{
\item Sean Matt and Ralph E. Pudritz: Physics \& Astronomy Department,
  McMaster University, Hamilton, Ontario L8S 4M1, Canada
  (\email{matt, pudritz@physics.mcmaster.ca}).}

\listofauthors{S. Matt \& R. E. Pudritz}
\indexauthor{Matt, S.}
\indexauthor{Pudritz, R. E.}

\shortauthor{Matt \& Pudritz}

\shorttitle{Disk Locking, Revisited}

\resumen{}

\abstract{In this talk, we take a new look at the theory of disk
locking, which assumes that an accreting protostar rids itself of
accreted angular momentum through a magnetic coupling with the
accretion disk.  We consider that differential rotation between the
star and disk twists the field lines.  For large enough twist, the
magnetic field lines connecting the star and disk open and disconnect.
This significantly reduces the spin-down torque on the star by the
disk, and so we find that disk-locking theory predicts spin periods
that are much too short to account for typical observed systems.}

\addkeyword{accretion, accretion discs}
\addkeyword{MHD}
\addkeyword{stars: formation}
\addkeyword{stars: magnetic fields}
\addkeyword{stars: pre-main-sequence}
\addkeyword{stars: rotation}

\begin{document}
\maketitle

\section{Introduction \label{sec_intro}}

The collapse of a molecular cloud naturally leads to a phase
consisting of a central protostar surrounded by a centrifugally
supported accretion disk \citep[for a review, see][]{bodenheimer95}.
Disk winds, MRI turbulence, and/or viscous processes remove angular
momentum from the disk and results in the accretion of material with
high specific angular momentum onto the star.  For typical parameters
for accreting protostars (CTTS's), the accretion alone will spin the
star up to near breakup speed in less than $\sim 10^5$ years, assuming
the star hadn't already formed at near breakup speed.  Since the
accretion lifetime is often greater than $10^6$ yr, the stars must rid
themselves of this excess angular momentum.  Further, it has been
generally accepted that accreting protostars spin more {\it slowly}
than their non-accreting counterparts \citep[e.g.,][]{bouvierea93}.
The general explanation is that the presence of an accretion disk
somehow regulates the stellar spin, and then after the disk is
dispersed, the star spins up as it contracts toward the main sequence.

\citet{konigl91} applied the neutron star accretion model of
\citet{ghoshlamb79} to accreting protostars and showed that a
``disk-locking'' (DL) mechanism could explain the coincidence of
accretion and slow rotation, in those systems.  According to DL
theory, magnetic field lines connect the star to the disk (acting as
``lever arms'') and carry torques that oppose and balance the angular
momentum deposited by accretion.  CTTS's are now known to posses
kilogauss-strength fields \citep[e.g.,][]{johnskrull3ea99}, and the
general DL model has been invoked by many authors.

Recent observations of CTTS's in Orion by \citet{stassunea99},
however, show no correlation between observed rotation periods and
accretion diagnostics, calling the standard DL scenario into question.
Furthermore, the magnetic fields of CTTS's, while strong, are
disordered \citep{safier98, johnskrullea99}, which reduces the
effectiveness of magnetic torques required for DL.  These developments
prompted us to revisit the general theory of DL.  In particular, the
connectivity between the star and disk is an important issue.  Much
recent work has shown that the magnetic connection between the star
and disk is severed when the magnetic field is highly twisted.  Here,
we show that the resulting spin-down torque is significantly reduced,
and the DL model cannot account for accreting stars that spin slowly
(e.g., $\sim$10\% of breakup speed).

\section{The ``Standard'' DL Model} \label{sec_standard}

\begin{figure}[!t]
\includegraphics[width=\columnwidth]{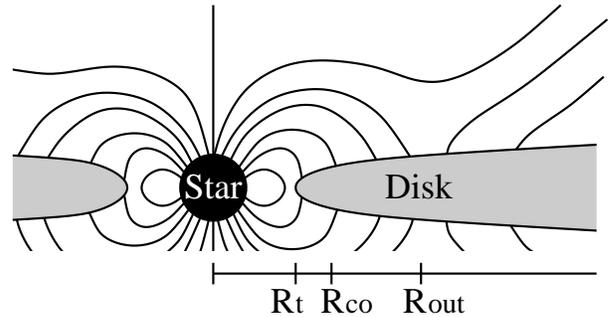}

\caption{Magnetic star-disk interaction.  \citep[from][]{mattpudritz04}
\label{fig_cartoon}}

\end{figure}

To begin, we first formulate a basic model that is representative of
the general DL picture discussed by many authors.  We follow the work
of \citet[][hereafter AC96]{armitageclarke96}.  The general theory
assumes the central star contains an axis-aligned dipolar magnetic
field.  A dipole is required because the field strength falls off as
slowly as nature allows ($r^{-3}$), and any higher order multipole
falls off so quickly that torques become negligible.  The dipole field
is anchored in the surface of the star and also connects to the
accretion disk, which is assumed to always be in Keplerian rotation.
The disk accretion rate $\mdotacc$ is constant in time and at all
radii in the disk.  Figure \ref{fig_cartoon} illustrates the basic
idea and identifies the location where the disk is truncated ($\rt$),
the outermost radius where the closed stellar field is connected to
the disk ($\rout$), and the corotation radius ($\rco$), where the
Keplerian angular rotation rate equals that of the star.  The usual
assumption is that $\rout \gg \rco$ (AC96 used $\rout \rightarrow
\infty$).

The magnetic torque on the star from field lines threading some range
of radii in the disk midplane is given by
\begin{eqnarray}
\label{eqn_torque}
\tmag = \int^{\rout}_{\rt} \gamma {\mu^2 \over r^4} \delta r
~~~~~~{\rm where}~~~~~~
\gamma \equiv {B_\phi \over B_z}
\end{eqnarray}
(e.g., AC96) where $\mu$ is the dipole moment, and $\gamma$ is the
``twist'' of the magnetic field.  So the torque depends not only on
the strength of the field but, more importantly, on how much it is
twisted.  The more it is twisted (larger $\gamma$), the stronger the
torques.

The field twist is generated by the differential rotation between the
star and disk.  As the field is twisted, it resists the twisting
(hence the torques) and slips backwards through the disk.  The larger
the $\gamma$, the faster the slipping.  When it can slip backward at
the same rate as the differential rotation rate between the star and
disk, a steady-state for $\gamma$ is achieved.  The speed of slipping
field lines is given by $v_{\rm d} = \beta v_{\rm kep} \gamma$, where
$\beta$ is a constant scale factor by which $v_{\rm d}$ compares to
the Keplerian speed, $v_{\rm kep}$ (AC96 use $\beta = 1$).  Thus, the
steady-state configuration of $\gamma(r)$ is given by $\gamma =
\beta^{-1} [(r/\rco)^{3/2} - 1]$, and so the torque in equation
\ref{eqn_torque} can be calculated.  The value of $\beta$ is unknown.
Standard $\alpha$-disk physics \citep{shakurasunyaev73} gives an upper
limit of $\beta \le 1$ and a likely value of a few orders of magnitude
lower.  We consider a value of $\beta = 10^{-2}$ as reasonable, but
given the uncertainties, we keep $\beta$ as a free parameter in our
analysis.  At first, we will use $\beta = 1$ (which gives the solution
of AC96), but we consider other values in the next section.

\begin{figure}[!t]
\includegraphics[bb=20 5 280 205,clip,width=\columnwidth]
  {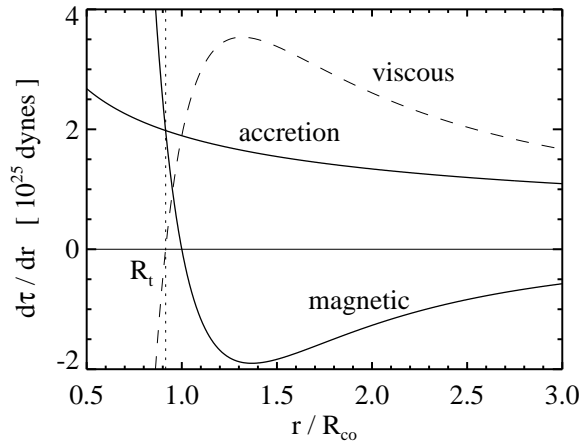}

\caption{Differential torques in the disk midplane for a system with
$\beta = 1$, $\mdotacc = 5 \times 10^{-8} M_\odot$yr$^{-1}$, $M_* =
M_\odot$, $R_* = 3 R_\odot$, $B_* = 10^3$ G, and a stellar spin period
of 5.7 days (so $\rco = 4.5 R_*$).  \citep[from][]{mattpudritz04} 
\label{fig_dtorques}}

\end{figure}

Figure \ref{fig_dtorques} shows the differential torques (per $\delta
r$) as a function of radius (normalized to $\rco$) in the disk
midplane, for a system with the parameters listed in the figure
caption.  The line labelled ``accretion'' represents torque that is
required to supply the steady-state $\mdotacc$.  The line labeled
``magnetic'' shows the differential torque (from eq.\
\ref{eqn_torque}) from the stellar field threading the disk.  Inside
$\rco$, this torque acts to spin the star up.  It decreases rapidly
away from the star as the dipole field becomes weaker.  It is zero at
$\rco$ because the differential rotation (and thus $\gamma$) is zero
there.  Outside $\rco$ it becomes stronger (though now spinning down
the star) as $\gamma$ increases, but it eventually becomes weaker
again because the decrease in the dipole field strength decreases
faster than the increase of the $\gamma$.  In order to satisfy the
steady-state condition, the disk must restructure itself so that the
sum of the magnetic and internal disk differential torques must equal
the accretion differential torque at all radii.  The dashed line
labelled ``viscous'' in Figure \ref{fig_dtorques} shows the required
internal disk torque.

The disk will be truncated near where the accretion and magnetic
differential torques are equal (and where viscous torque = 0).  From
that point ($\rt$) inward, all of the specific angular momentum of the
disk material will end up on the star.  So, to calculate the net
torque on the star from the accretion of disk material, $\tacc$, one
integrates the differential accretion torque (shown in Fig.\
\ref{fig_dtorques}) from $\rt$ to the surface of the star.  The net
magnetic torque, $\tmag$, is obtained by integrating equation
\ref{eqn_torque} from $\rt$ to $\rout$ (which is thus far assumed to
be $\infty$).

For any given values of $M_*$, $R_*$, $B_*$, $\mdotacc$, and the
stellar rotation period, this ``standard'' theory gives the net torque
on the star.  The system is stable in that, for fast rotation, the net
torque spins the star down, and for slow rotation, the star spins up.
Also, for typical CTTS parameters, the torques spin the star up or
down in $\sim 10^5$ yr, so one expects that most systems will exist in
a spin equilibrium state where the net torque on the star is zero.
The ``standard'' DL model thus predicts the spin period in the
equilibrium state, $\spineq$, at which the system is ``disk locked.''
Figure \ref{fig_dtorques} is shown in its equilibrium spin state
($\spineq = 5.7$ d).

Models such as this have had success at explaining the spin rates of
slow rotators.  For example, the well-studied CTTS, BP Tau has
$\mdotacc = 3 \times 10^{-8} M_\odot$ yr$^{-1}$, $R_* = 2 R_\odot$,
and $M_* = 0.5 M_\odot$ \citep{gullbringea98}.  Using the mean field
strength of 2 kG found by \citet{johnskrull3ea99}, our ``standard'' DL
theory predicts $\spineq = 7.5$ d (corresponding to 6\% of breakup
speed)---remarkably similar to the observed value of 7.6 d
\citep{vrbaea86}.  Thus the DL theory {\it seems} to work, but there's
at least one major problem.  Namely, we assumed that the star and disk
were connected to $\rout \rightarrow \infty$.  At large radii, the
field will be highly twisted, and there is a physical limit to that
twist, which we have so far ignored.  In the next section, we consider
the effect on the DL model of an upper limit to the magnetic twist.

\section{Effect of Limited Twist}

\begin{figure}[!t]
\includegraphics[bb=20 5 280 205,clip,width=\columnwidth]
  {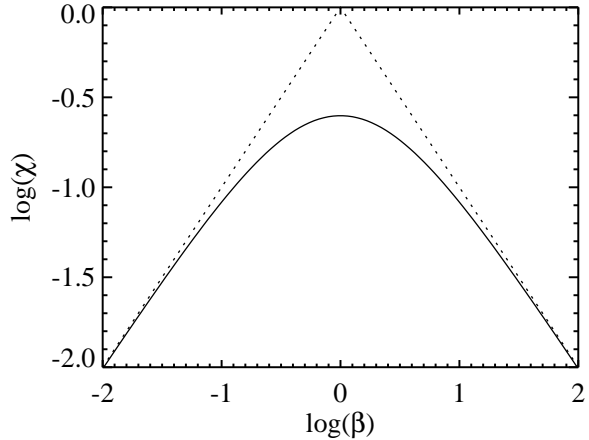}

\caption{Logarithm of the net magnetic torque as a function of
$\log(\beta)$.  The torque (denoted $\chi$, see text) is normalized to
the AC96 value, so that the value of $\chi = 1$ is the torque from the
``standard'' DL theory, and the solid line shows the effect of limited
magnetic twist.  For reference, the dotted lines show $\chi = \beta$
and $\chi = \beta^{-1}$.  \citep[from][]{mattpudritz04}
\label{fig_chi}}
\end{figure}

Twisted magnetic fields that connect the star to the disk exert
torques between the two.  The larger the region in the disk that is
magnetically connected to the star, the larger is the total magnetic
torque.  So the actual location of $\rout$ is important, since it
delimits the connected region (i.e., it determines the integration
range in eq.\ \ref{eqn_torque}).  Many recent studies \citep[see][and
references therein]{uzdensky3ea02} have shown that dipole magnetic
field lines connected to a rotating disk open up, when twisted past a
critical value of $\gamma = \gammacrit \approx 1$.  This is
unavoidable and occurs because the magnetic pressure associated with
the azimuthal component of magnetic field pushes out against the
poloidal field lines.  These open field lines (see, e.g., the field
lines outside $\rout$ in Fig.\ \ref{fig_cartoon}) cannot convey
torques between star and disk.

Since the steady-state value of $\gamma$ increases as $r^{3/2}$ away
from $\rco$ (see \S \ref{sec_standard}), it inevitably reaches the
critical value, which we take as $\gammacrit = 1$.  As an
approximation, we assume that the star is connected as before to the
disk, but that the extent of connected region is now limited to the
region where $\gamma \le \gammacrit$.  The outermost location of this
region is $\rout= (1 + \beta \gammacrit)^{2/3} \rco$, beyond which, we
assume the star is disconnected from the disk, and so the differential
torque is zero.  Since the size of the connected region is smaller
than for the ``standard'' model, the net magnetic torque is less.

Figure \ref{fig_chi} illustrates how the magnetic torque is altered by
the opening of field lines, as a function of the parameter $\beta$.
Shown is the value of the net magnetic torque (denoted $\chi$),
renormalized so that the torque predicted by the ``standard'' DL
theory of the previous section ($\beta = 1$ and $\rout = \infty$)
gives a value of $\chi = 1$.  The normalization allows Figure
\ref{fig_chi} to be valid for any given values of $\mdotacc$, $M_*$,
$\mu$, and the stellar spin period.  It is evident that the magnetic
torque will always (for any $\beta$) be significantly less than
predicted by the ``standard'' theory.  The dependence of $\chi$ on
$\beta$ can be understood as a competition between two different
effects: One is that $\rout$ decreases for decreasing $\beta$,
reducing the integration range of equation \ref{eqn_torque}; the other
is that the steady-state $\gamma$ decreases for increasing $\beta$,
reducing the differential torque at each radius.  For the critical
value of $\beta = 1$, these two effects conspire to give a maximal
value of $\tmag$ that is four times less than predicted by the
``standard'' DL theory of section \ref{sec_standard} (for any given
values of $\mdotacc$, $M_*$, $\Omega_*$, and $\mu$).  So by using
$\beta = 1$ above, we have considered the ``best possible case'' for
DL theory, since $\tmag$ is less for all other values of $\beta$.  A
reduced magnetic torque, means that the star must spin faster before
it is in equilibrium.  A faster spin reduces $\rco$, so the torques
come from closer to the star where the dipole field is stronger,
making up for the decreased integration range.

We can now revisit our example case of BP Tau.  Using the ``updated''
DL theory ($\gammacrit = 1$), the ``best case'' value of $\beta = 1$
predicts $\spineq = 4.1$ d.  For $\beta = 0.1$ (or $\beta = 10$),
$\spineq = 2.5$ d.  The time to spin up from 7.6 d to 4.1 d (or even
to 2.5 d) is $1 \times 10^5$ yr, which is significantly shorter than
BP Tau's age of $6 \times 10^5$ yr \citep{gullbringea98}.  Therefore,
BP Tau has either gone through a recent change in (e.g.)\ $\mdotacc$,
so that it is not currently in the equilibrium state, or the model is
incomplete.  In order for BP Tau to currently be in an equilibrium
spin state, there must be significant spin-down torques on the star
other than the torques along field lines connecting it to the disk.
As an aside, we note that we have thus far used the mean stellar
magnetic field strength of 2 kG found by \citet{johnskrullea99},
though this is not the true strength of the {\it dipole} field.  If
instead we use the $3 \sigma$ upper limit to the dipole field strength
of 200 G \citep{johnskrull3ea99}, even the ``standard'' theory
predicts $\spineq = 1.0$ d.

\section{Summary of Problems With DL}

The disk-locking scenario has recently been called into question by
observations, as well as by theoretical considerations.  In
particular, there are four, completely independent issues:
\begin{enumerate}

\item{\citet{stassunea99} found no correlation between accretion
parameters and spin rates of TTS in Orion.}

\item{CTTS's apparently do not have strong {\it dipole} fields
\citep[e.g.,][]{safier98, johnskrullea99}.}

\item{Stellar winds are expected to open field lines that would
otherwise connect to the disk \citep{safier98}.  A disk wind would
have a similar effect.}

\item{Finally, a large portion of the magnetic field connecting the
star to the disk will open up, due to the differential rotation
between the two \citep[e.g.,][]{uzdensky3ea02}.  We have shown that
the resulting spin-down torque on the star by the disk is less (by at
least a factor of four and possibly by orders of magnitude) than
calculated by previous authors.  The predicted equilibrium spin rate
is therefore much faster.}

\end{enumerate}
\adjustfinalcols
So the DL scenario does not explain the angular momentum loss of the
so called slow rotators---the group originally targeted by DL theory.

We conclude that, in order for accreting protostars to spin as slowly
as 10\% of breakup speed, there must be spin-down torques acting on
the star other than those carried by magnetic field lines connecting
the star to the disk.  The presence of open stellar field lines
naturally leads to the possibility that excess angular momentum is
carried by a stellar wind along those open lines.  We plan to
investigate this possibility in the near future.



\acknowledgements

We'd like to thank the organizers for a fantastic meeting.  This
research was supported by NSERC, McMaster University, and CITA,
through a CITA National Fellowship to S. M.






\end{document}